\documentclass[twocolumn,showpacs,prc,superscriptaddress,amsmath,amssymb]{revtex4-1}

\usepackage{multirow}
\usepackage{graphicx}
\usepackage{dcolumn}
\usepackage{bm}
\usepackage{soul}
\usepackage{color}

\begin{document}

\preprint{APS/123-QED}

\title{Sub-barrier fusion enhancement with radioactive $^{134}$Te}

\author{Z. Kohley}
 \email{kohley@kohley@nscl.msu.edu}
\affiliation{National Superconducting Cyclotron Laboratory, Michigan State University, East Lansing, Michigan 48824, USA}
\affiliation{Chemistry Department, Michigan State University, East Lansing, Michigan 48824, USA}
\affiliation{Physics Division, Oak Ridge National Laboratory, Oak Ridge, TN 37831, USA}
\author{J. F. Liang}
\author{D. Shapira}
\author{C.~J. Gross}
\author{R. L. Varner}
\affiliation{Physics Division, Oak Ridge National Laboratory, Oak Ridge, TN 37831, USA}
\author{J. M. Allmond}
\affiliation{Joint Institute for Heavy Ion Research, Oak Ridge National Laboratory, Oak Ridge, TN 37831, USA}
\author{J. J. Kolata}
\affiliation{Physics Department, University of Notre Dame, Indiana 46556, USA}
\author{P. E. Mueller}
\affiliation{Physics Division, Oak Ridge National Laboratory, Oak Ridge, TN 37831, USA}
\author{A. Roberts}
\affiliation{Physics Department, University of Notre Dame, Indiana 46556, USA}

\date{\today}

\begin{abstract}
The fusion cross sections of radioactive $^{134}$Te~+~$^{40}$Ca were measured at energies above and below the Coulomb barrier.  The evaporation residues produced in the reaction were detected in a zero-degree ionization chamber providing high efficiency for inverse kinematics.  Both coupled-channel calculations and comparison with similar Sn+Ca systems indicate an increased sub-barrier fusion probability that is correlated with the presence of positive Q-value neutron transfer channels.  In comparison, the measured fusion excitation functions of $^{130}$Te~+~$^{58,64}$Ni, which have positive Q-value neutron transfer channels, were accurately reproduced by coupled-channel calculations including only inelastic excitations.  The results demonstrate that the coupling of transfer channels can lead to enhanced sub-barrier fusion but this is not directly correlated with positive Q-value neutron transfer channels in all cases.
\end{abstract}


\maketitle
\section{Introduction}
Heavy-ion fusion below the Coulomb barrier presents the opportunity to study the quantum tunneling of many-body systems~\cite{DASGUPTA98,BALAN98}.  For relatively light systems ($Z_{1}Z_{2}~\lesssim~250)$, the fusion probability below the barrier can be described well using the single barrier penetration model (BPM)~\cite{BALAN98,Vaz81}.  However, for heavier systems dramatic enhancements beyond the BPM were observed in the fusion probability below the barrier~\cite{DASGUPTA98,BALAN98,Vaz81,REISDORF94}.  This was discovered to be associated with the coupling of additional degrees of freedom, such as vibrational and rotational excitations, which modify the single Coulomb barrier producing a distribution of barriers~\cite{DASGUPTA98,BALAN98}.  Coupled-channel calculations provide a powerful tool to calculate the fusion excitation functions with the inclusion of couplings to inelastic excitations and/or rotational bands~\cite{BALAN98,Hag12}.

\par
While nucleon transfer has been considered a ``doorway'' to heavy-ion fusion~\cite{Hen87}, the coupling of transfer channels to the sub-barrier fusion process is not fully understood~\cite{DASGUPTA98,Hag12,Ste88,Koh11,Jia12}.  A variety of experimental measurements have shown strong correlations between the presence of positive Q-value neutron transfer (PQNT) channels and enhanced sub-barrier fusion probabilities (see Refs.~\cite{DASGUPTA98,Bor92,STEF95,STEF07,STEF06,TIMMERS98,TROTTA01,SCARL00,KALKAL10,ZHANG10,Kol12,Mont13}  and references therein).  For example, the fusion of the symmetric $^{40}$Ca+$^{40}$Ca and $^{48}$Ca+$^{48}$Ca systems show very similar excitation functions when taking into account the differences in the barrier height and nuclear radii. However, the fusion probability is over an order of magnitude greater for the cross system, $^{40}$Ca+$^{48}$Ca, at $E_{c.m.}/V_{C}\sim 0.94$ (where $V_{C}$ is the Coulomb barrier)~\cite{TROTTA01}.  The most significant difference of the cross system compared to the symmetric systems is the presence of PQNT channels, which suggests that the enhanced sub-barrier fusion is related to the coupling to transfer channels.  In contrast, we previously examined the fusion excitation functions of different Sn+Ni and Te+Ni systems, which have large variations in the Q-value and number of PQNT channels, and observed no significant enhancement in relation to the PQNT channels~\cite{Koh11}.  A recent measurement of $^{16,18}$O~+~$^{76,74}$Ge fusion by Jia \textit{et al.} showed a similar lack of enhancement with respect to the presence of PQNT channels~\cite{Jia12}. Thus, it is unclear why the enhancement from couplings to transfer channels is only present in certain systems.

\par
Presently, it is difficult to included transfer couplings in a consistent manner into the coupled-channel calculations~\cite{REISDORF94,DASGUPTA98,BALAN98,CANTO06,DENISOV00,Hag12,ESBENSEN98,Mont13}.  Zagrebaev proposed a semi-empirical formulation to account for the coupling of transfer channels with positive Q-values which demonstrated reasonable agreement with both fusion excitation functions and neutron transfer cross sections~\cite{ZAGREB03}.  Using a quantum diffusion approach, Sargsyan \textit{et al.} proposed that the fusion enhancement could be attributed to changes in the deformations of the nuclei from nucleon transfer~\cite{Sar12}.  Recently Oberacker and Umar addressed the issue using a microscopic approach with the density constrained time-dependent Hartee-Fock (DC-TDHF) method and examined the fusion of $^{132,124}$Sn~+~$^{40,48}$Ca~\cite{Ober13}.  The DC-TDHF results showed that the enhancement in the fusion of the $^{132}$Sn~+~$^{40}$Ca system (which has many PQNT channels) was due to a narrower width of the ion-ion potential for that system.  Systematic comparisons between different measured fusion excitation functions~\cite{STEF07,Koh11,Kol12} and continued theoretical efforts should help in understanding the mechanism in which transfer channels effect the fusion process.  In the present work, the fusion excitation function of $^{134}$Te~+~$^{40}$Ca was measured to provide additional insight into the role of PQNT channels through comparison with the $^{132,124}$Sn~+~$^{40,48}$Ca fusion measurements of Kolata \emph{et al.}~\cite{Kol12} and coupled-channel calculations.


\section{Experimental Setup and Analysis}
\par
The radioactive $^{134}$Te beam was produced at the Holifield Radioactive Ion Beam Facility (HRIBF) using the Isotope Separation Online (ISOL) technique.  A 10-15 $\mu A$ proton beam produced by the Oak Ridge Isochronous Cyclotron bombarded a uranium carbide target.  The fragments produced from the proton-induced fission of the uranium were ionized using a plasma ionization source. The $^{134}$Te fragments were selected using mass separators and passed to the 25~MV tandem electrostatic accelerator, which accelerated the beam to energies between 535 and 619 MeV. Further information on the HRIBF ISOL technique and radioactive ion beam production procedures can be found in Ref.~\cite{HRIBF}.  Beam rates for the $^{134}$Te ranged from 70,000 to 90,000 particles per second.  The $^{134}$Te beam impinged on a 0.466 mg/cm$^{2}$ $^{40}$CaF$_{2}$ target.  As discussed in Ref.~\cite{Kol12}, the target composition and thickness were determined using the Rutherford Backscattering Spectrometry (RBS) at Hope College in Holland, MI.

\par
The evaporation residues (ERs) produced in the fusion of Te~+~Ca (and Te~+~F) were detected in the Compact Setup for Studies of Evaporation Residues (CSSER)~\cite{SHAPRIA05}.  The detector system was designed to provide high efficiency measurements of evaporation residues produced through inverse kinematics.  The setup consisted of two microchannel plate timing detectors (MCPs) placed upstream from the target that monitored the beam rate and served as a timing references.  Downstream from the target, a position sensitive MCP was mounted followed by a zero-degree ionization chamber.  The configuration was identical to that used in Ref.~\cite{Kol12}, where the distance between the target and third MCP was increased from 169~mm (standard configuration~\cite{SHAPRIA05,LIANG07,Koh11,Lia12}) to 329~mm to increase the time of flight (ToF) path length.

\par
\begin{figure}
\includegraphics[width=0.47\textwidth]{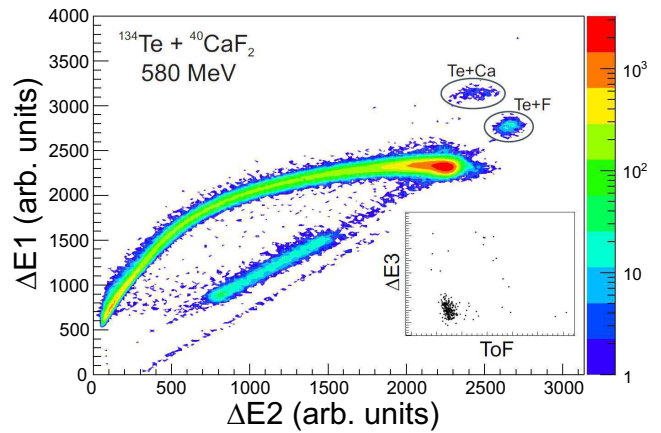}
\caption{\label{f:er} (color online) $\Delta$E1-$\Delta$E2 contour plot from the ionization chamber for ER triggered events from the $^{134}$Te+$^{40}$Ca reaction.  The two groups of ERs from the Te~+~Ca and Te~+~F fusion are indicated and are clearly separated from the beam-like particles.  The insert shows the $\Delta$E3-ToF for the events selected from the Te~+~Ca ER gate.}
\end{figure}

Since the ERs will travel at a reduced velocity relative to the beam particles, events with an increased ToF between the second and third MCPs were selected for identifying ERs.  The ERs were identified from the energy-loss signals in the ionization chamber.  The ionization chamber has three anodes providing $\Delta$E1, $\Delta$E2 and $\Delta$E3 signals.  The chamber was filled with CF$_{4}$ gas at a pressure of 33~Torr. Fig.~\ref{f:er} shows the $\Delta$E1-$\Delta$E2 plot for ER triggered events for the $^{134}$Te+$^{40}$Ca reaction at 580~MeV.  The ERs from the fusion of both Te~+~Ca and Te~+~F are clearly discernible and are well separated from the beam-like particles.  A 2-D gate was drawn to select the ERs.  Background events were removed by examining the ToF-$\Delta$E3 histogram of the gated ER events, as shown in the insert of Fig.~\ref{f:er}.  The systematic error was estimated to be at most 5$\%$ due to reasonable variations in the ER selection gates.

\par
The efficiency of the ionization chamber at each energy was estimated using the $\texttt{\sc pace2}$ statistical model~\cite{PACE2} to calculate the angular distributions of ERs.  The energy loss and straggling effects in the target and MCPs were taken into account in the Monte Carlo simulation.  The strong forward focus of the ERs, owing to the inverse kinematics, in combination with the zero-degree ionization chamber resulted in high detection efficiencies of about 97$\%$ for all energies. The beam, downscaled by 1000, was also counted in the ionization chamber allowing for the ER cross sections to be calculated. Since the energy loss of the beam in the target is about 30~MeV, the effective reaction energy was determined by the fitted spline method~\cite{LIANG07}.  The effective reaction energy will be between the initial beam energy (E$_{in}$) and the energy after passing through the target (E$_{out}$). Below the barrier, where the fusion cross section changes rapidly, the effective reaction energy will be weighted towards E$_{in}$, as this corresponds to a much higher cross section.  The fitted spline method has been tested and shown to provide accurate determination of the effective reaction energies in thick target measurements~\cite{LIANG07}.  The SRIM energy loss code~\cite{SRIM} and spline fitting classes from the ROOT software package~\cite{ROOT} were used to implement the method.

\section{Results and Discussion}

\begin{figure}
\includegraphics[width=0.40\textwidth]{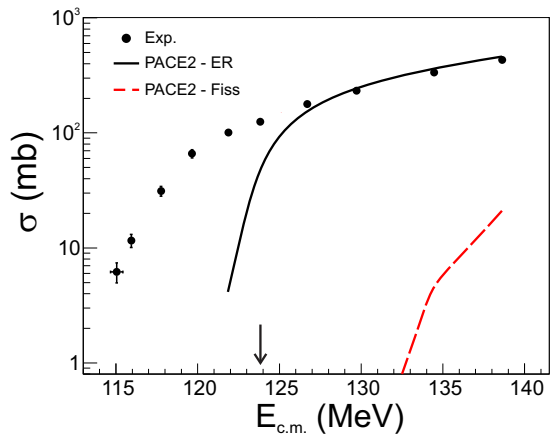}
\caption{\label{f:pace} (color online) The ER cross sections as a function of the center of mass energy from the $^{134}$Te~+~$^{40}$Ca system.  Calculations of the ER and fission cross sections from $\texttt{\sc pace2}$ are shown for comparison as the solid and dashed lines, respectively.}
\end{figure}

The evaporation residue cross sections as a function of the center of mass energy from the $^{134}$Te~+~$^{40}$Ca reaction are presented in Fig.~\ref{f:pace}.  The experimental results are compared to ER and fission cross sections calculated from the statistical model $\texttt{\sc pace2}$~\cite{PACE2}, shown as the solid and dashed lines.  The fusion cross sections in $\texttt{\sc pace2}$ are taken from the Bass systematics~\cite{BASS74}.  The calculation was completed using a level density parameter of A/8 MeV$^{-1}$, saddle point to ground state level density ratio ($a_{f}/a_{n}$) of 1.04, spin distribution diffuseness of 4$\hbar$, Sierk fission barriers, and experimentally determined masses.  This choice of input parameters had been shown to provide good agreement between the $\texttt{\sc pace2}$ and experimental cross sections for a variety of Sn~+~Ni reaction systems~\cite{LIANG07}.

\par
The ER cross sections from the $\texttt{\sc pace2}$ calculations agree well with the experimental data above the Coulomb barrier, which is indicated by the arrow.  Deviations between $\texttt{\sc pace2}$ and the experimental data around the Coulomb barrier are expected since the Bass systematics do not extend below the barrier.  The $\texttt{\sc pace2}$ calculations show that the measured ER cross sections can be safely approximated to be equivalent to the total fusion cross section.  Even at the highest energy the fission cross section is calculated to be over one order of magnitude smaller than the ER cross section.

\par
\begin{figure}
\includegraphics[width=0.43\textwidth]{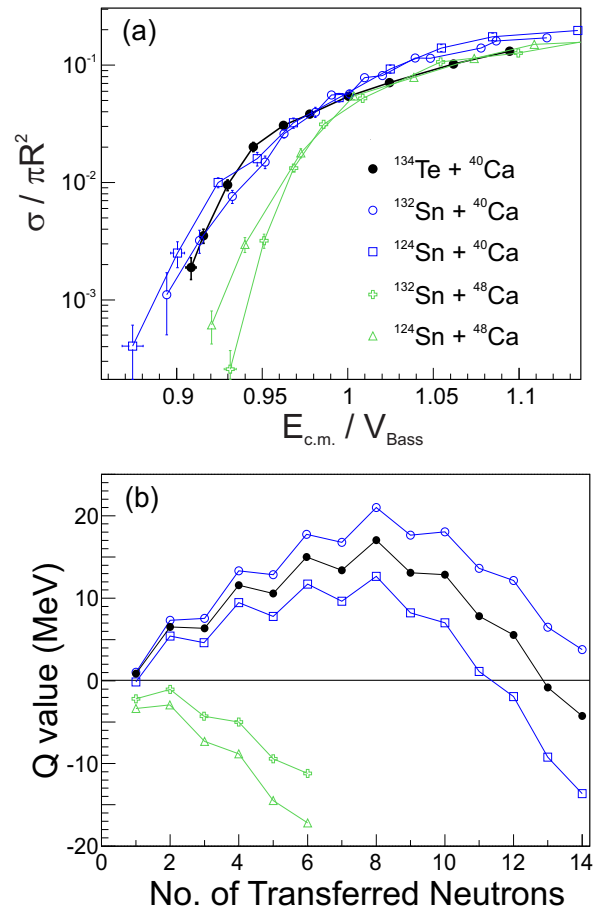}
\caption{\label{f:snni} (color online) (a) $^{134}$Te~+~$^{40}$Ca reduced excitation function is compared to the $^{132,124}$Sn~+~$^{40,48}$Ca systems from Ref.~\cite{Kol12}.  (b) Neutron transfer Q-values, calculated from the ground state masses, as a function of the number of neutrons transferred from the projectile to the target.}
\end{figure}

The reduced fusion excitation functions from the $^{134}$Te~+~$^{40}$Ca system and similar $^{132,124}$Sn~+~$^{40,48}$Ca systems~\cite{Kol12} are compared in Fig.~\ref{f:snni}(a) to gain insight into the influence of transfer channels on the sub-barrier enhancement.  The cross section was scaled by $\pi R^{2}$, where $R = 1.2 (A_{proj}^{1/3}+A_{tgt}^{1/3})$~fm, to remove the differences in the nuclear radii of the systems.  The center of mass energy was scaled by the Bass fusion barrier ($V_{Bass}$)~\cite{BASS74} to account for systematic changes in the barrier position.
\par
The reduced fusion excitation functions are clearly separated into two groups according to the $^{40}$Ca and $^{48}$Ca targets.  There is a significant enhancement in the sub-barrier fusion for the reactions with $^{40}$Ca in comparison to $^{48}$Ca.  The enhancement is correlated with the presence of PQNT channels as shown in Fig.~\ref{f:snni}(b).  The $^{132,124}$Sn~+~$^{48}$Ca reactions do not have any PQNT channels, whereas the $^{40}$Ca systems each have at least 10 PQNT channels.    The results suggest that the influence of the transfer channels is similar in the Te~+~Ca and Sn~+~Ca reactions.

An issue to be discussed in the systematic comparison of the reduced excitation functions [Fig.~\ref{f:snni}(a)] is the strength of the octupole mode in $^{40}$Ca which can produce a potential renormalization of the barrier~\cite{Hag97}.  This was demonstrated by Stefanini \emph{et al.} for Ca~+~Zr systems~\cite{STEF07}.  The potential renormalization produces an increase in the above barrier fusion cross section for the $^{40}$Ca systems in comparison to the $^{48}$Ca systems.  Stefanini \emph{et al.} proposed a method of shifting the different excitation functions by a constant energy such that the above barrier cross sections are equivalent.  This should remove the differences due to the renormalization of the potential~\cite{STEF07}.  The method was applied to the Sn~+~Ca systems and the shifted excitation functions still presented an enhancement for the $^{40}$Ca systems.  This suggests that the enhancement of the $^{40}$Ca systems is not solely due to the strong octupole mode but appears to be related to the presence of PQNT channels.  However, fluctuations in the above barrier cross sections of the Sn + Ca systems, likely due to the difficulty of the RIB experiment, create some uncertainty in shift constant.  Therefore, it is possible that a larger shift constant could be required which would reduce the enhancement of the $^{40}$Ca systems.

\par
While the comparison with the Sn~+~Ca systems indicated an enhancement due to the transfer channel couplings, it is important to examine if couplings to inelastic excitations in the $^{134}$Te~+~$^{40}$Ca fusion can reproduce the sub-barrier enhancement.  The coupled-channels code $\texttt{\sc ccfull}$~\cite{CCFULL}, which includes both linear and non-linear couplings, was used to solve the coupled-channel equations.  $\texttt{\sc ccfull}$ uses the incoming wave boundary condition, which requires a deep potential for calculating fusion cross sections at high energies~\cite{NEWTON04}.  Therefore, the diffuseness ($a_{0}$) and radius ($r_{0}$) parameters from a potential with a depth ($V_{0}$) of 115~MeV were varied to match the shape of the barrier calculated from the Aky\"{u}z-Winther (AW) potential~\cite{AW81}.  A comparison of the AW and modified deep potential are shown in Table~\ref{t:pot}.   The modified deep potential was used in the subsequent $\texttt{\sc ccfull}$ calculations.


\begin{table*}
\begin{center}
\caption{A comparison of the AW and modified deep potential and associated barrier characteristics for the $^{134}$Te~+~$^{40}$Ca system.}
\begin{tabular}{m{2cm} m{2cm} m{2cm} m{2cm} m{2cm} m{2cm} m{2cm}}
\hline
\hline
Potential &$V_{0}$ (MeV)  &$a_{0}$ (fm)    &$r_{0}$ (fm)  &$V_{B}$ (MeV) &$\hbar\omega$ (MeV) &$R_{B}$ (fm) \\
\hline
AW        &78.26       &0.679            &1.18              &121.8         &3.87                &11.48\\
Mod. Deep  &115       &0.756            &1.12              &121.6         &3.80                &11.42\\
\hline
\hline
\end{tabular}
\label{t:pot}
\end{center}
\end{table*}

\par
\begin{figure}
\includegraphics[width=0.40\textwidth]{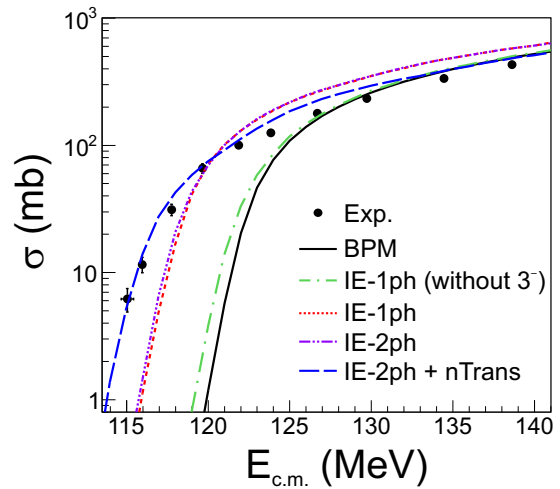}
  \caption{\label{f:cc} (color online) $\texttt{\sc ccfull}$ calculations for the fusion of $^{134}$Te~+~$^{40}$Ca with no couplings (BPM), inelastic one-phonon (1ph) and two-phonon (2ph) excitation (IE) couplings, and IE with neutron transfer couplings.  The experimental data are shown as the solid black circles.}
\end{figure}

The $^{134}$Te~+~$^{40}$Ca fusion excitation function is compared to the BPM in Fig.~\ref{f:cc}.  While the calculation and data agree above the barrier, the BPM significantly under predicts the sub-barrier fusion probability.  Therefore, the one-phonon excitations (1ph) of the $2^{+}$ and $3^{-}$ states of $^{40}$Ca and the $2^{+}$ state of $^{134}$Te were included into the coupling scheme.  The $^{134}$Te $3^{-}$ state, recently measured at 3749~keV~\cite{All12}, was not included in the calculation since the $B(E3)$ is unknown.  The quadruple and octupole deformations of $^{40}$Ca were calculated from the $B(E2)$ and $B(E3)$ reduced transition probabilities from Refs.~\cite{BE2} and \cite{BE3}.  The $B(E2)$ value for $^{134}$Te was taken from the HRIBF Coulomb excitation measurements~\cite{Rad02,Dan11}.  The $\texttt{\sc ccfull}$ calculation including inelastic couplings (IE), shown as the red dashed line in Fig.~\ref{f:cc}, increases the sub-barrier fusion enhancement but is still unable to reproduce the experimental data.
\par
As mentioned previously, the increase in the IE coupled fusion cross sections above the barrier are related to the strong octupole state in $^{40}$Ca at 3.7~MeV, which can produce a static potential renormalization.  The curvature of the barrier ($\hbar\omega$) defines a limit for which states need to be included in the coupled-channel calculations~\cite{Hag97}.   Since the energy of the $3^{-}$ state in $^{40}$Ca is roughly equal to the curvature of the barrier it was included in the coupling scheme~\cite{Hag97}. The removal of the $3^{-}$ state from the calculation is shown as the green dot-dashed line in Fig.~\ref{f:cc}.  Without the $3^{-}$ state, the above barrier cross sections are very similar to the BPM and only a slight increase in the sub-barrier fusion is observed with respect to the BPM.  In either case (including or excluding the $3^{-}$ state), the general conclusion that the measured fusion excitation function cannot be reproduced through inclusion of only one-phonon IE couplings remains the same.

\par
Since the coupling of the one-phonon states was unable to reproduce the experimental data, two-phonon (2ph) couplings were included in the calculations and are shown in Fig.~\ref{f:cc}.  The two-phonon (2$^{+}$)$^{2}$ and (3$^{-}$)$^{2}$ states were included along with all cross-coupling terms, such as $2^{+} \otimes 3^{-}$.  Multi-phonon couplings have previously been shown to be important for reproducing the barrier distributions of certain systems~\cite{Row10,STEF95,Stefprl95,Esb05}.  The inclusion of the two-phonon excitations slightly increases the sub-barrier fusion but the large discrepancy between the coupled-channel calculations and experimental data remains.

\begin{figure}
\includegraphics[width=0.35\textwidth]{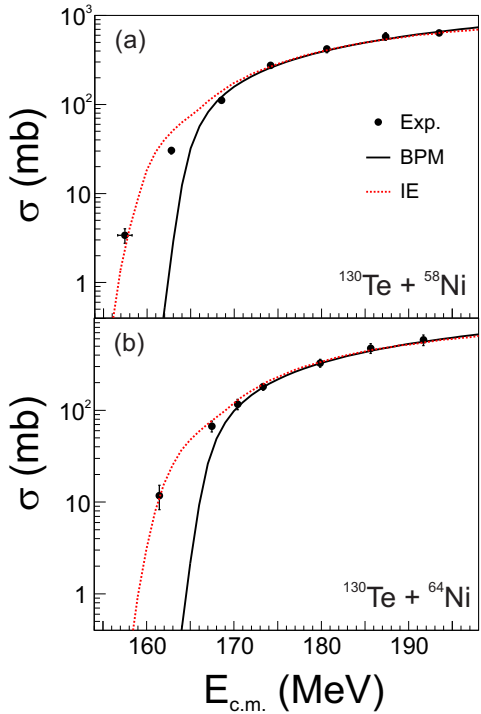}
\caption{\label{f:teni} (color online) Measured fusion excitation functions of (a) $^{130}$Te~+~$^{58}$Ni and (b) $^{130}$Te~+~$^{64}$Ni compared to coupled-channel calculations without (BPM) and with inelastic excitation (IE) channel couplings.}
\end{figure}

\par
In order to reproduce the sub-barrier fusion enhancement coupling to the transfer channels is required.  Therefore, the coupling of the $1n$ transfer channel from $^{134}$Te to $^{40}$Ca was included with a Q-value of 0.86~MeV.  Since experimental transfer measurements do not exist for this system the transfer form factor is unknown and only a qualitative measure of the transfer coupling can be extracted. Thus, the coupling constant ($F_{t}$) in the $\texttt{\sc ccfull}$ calculations was varied until reasonable agreement with experimental data was achieved, which corresponded to $F_{t}$~=~0.55.  As shown in Fig.~\ref{f:cc}, the coupled-channel calculations with both the IE and $1n$ transfer channels are able to reproduce the sub-barrier fusion enhancement.  Together, the results of the coupled-channel calculations and comparisons with the Sn~+~Ca systems demonstrate that the fusion of $^{134}$Te~+~$^{40}$Ca is clearly modified by the presence of the PQNT~~\cite{Koh11}.

\par
The role of transfer channels in the $^{134}$Te~+~$^{40}$Ca system is different than that in the previously studied $^{130}$Te~+~$^{58,64}$Ni systems~\cite{Koh11}.  While the structure of $^{134}$Te and $^{130}$Te are similar, the $^{40}$Ca and $^{58,64}$Ni targets are very different. In comparison to the Ni targets, the doubly magic $^{40}$Ca has a much higher lying $2^{+}$ state and, as discussed above, a strong coupling of the $3^{-}$ state. Even though the $^{130}$Te~+~$^{58}$Ni system has 11 PQNT channels the reduced excitation function showed no enhancement beyond the $^{130}$Te~+~$^{64}$Ni system, which has only 1 PQNT channel.

\par
In our previous work~\cite{Koh11} the coupled-channel calculations for the $^{130}$Te~+~$^{58,64}$Ni systems were not presented.  In Fig.~\ref{f:teni} the fusion excitation functions calculated using $\texttt{\sc ccfull}$ with no couplings and IE couplings are shown.  The calculations were completed identically to that described above for the $^{134}$Te~+~$^{40}$Ca system.  The $2^{+}$ and $3^{-}$ states of both the projectile and target were included in the IE coupling scheme~\cite{BE2,BE3}.  The experimental excitation functions are reproduced well by the coupled-channel calculations including only IE channels.  A clear discrepancy is observed pertaining to the role of transfer channels in the fusion process.  Even though the $^{134}$Te~+~$^{40}$Ca and $^{130}$Te~+~$^{58}$Ni systems have similar numbers of PQNT channels (12 and 11, respectively) the coupled-channel calculations show that the $^{130}$Te~+~$^{58}$Ni data can be explained by IE couplings while the $^{134}$Te~+~$^{40}$Ca data require the inclusion of transfer channel couplings.  This suggests that the unique structure of the $^{40}$Ca target is associated with producing stronger transfer channel effects.

\section{Summary}
\par
The fusion cross sections of $^{134}$Te~+~$^{40}$Ca were measured at energies below and above the Coulomb barrier using the CSSER setup at the HRIBF to directly detect the evaporation residues.  A large number of PQNT channels result from the extreme difference in the neutron richness of the radioactive $^{134}$Te and stable $^{40}$Ca.  Comparing the reduced excitation function of the $^{134}$Te~+~$^{40}$Ca system to the $^{132,124}$Sn~+~$^{40,48}$Ca systems showed a systematic increase in the sub-barrier fusion cross sections correlated with the presence of PQNT channels.  Coupled-channel calculations also required the inclusion of transfer channel couplings, along with inelastic excitations, to reproduce the measured $^{134}$Te~+~$^{40}$Ca excitation function. In comparison, the fusion excitation functions of the $^{130}$Te~+~$^{58}$Ni and $^{130}$Te~+~$^{64}$Ni systems were well reproduced by coupled-channel calculations including only inelastic excitations despite having 11 and 1 PQNT channels, respectively.  Thus, it appears that the structure of $^{40}$Ca, in comparison to the $^{58,64}$Ni, leads to more pronounced transfer channel effects~\cite{Liang13}.  The role of transfer channels in the fusion process is complex and the results indicate that presence of PQNT channels does not necessitate additional fusion enhancement. Additional experimental and theoretical effort is needed to begin to unravel the relationship between transfer and fusion reactions.  Specifically, experimental measurements of \emph{both} the fusion and multinucleon transfer cross sections, such as the work of completed at Argonne National Laboratory for the Sn~+~Ni systems~\cite{FREEMAN83,LESKO86,WOLF87,ESBENSEN98,JIANG98} and the Ca~+~Zr measurements at INFN~\cite{TIMMERS98,STEF07,Szi07,Corr11}, are strongly encouraged as this can help constrain the theoretical calculations and provide a deeper understanding of the coupling of the fusion and transfer processes.


\section{Acknowledgements}
We thank Dr. K. Hagino for the providing the $\texttt{\sc ccfull}$ code.  We would also like to thank the staff members of the Holifield Radioactive Ion Beam Facility for the excellent quality radioactive and stable beams.  This research was supported by the DOE Office of Nuclear Physics and NSF Grant Nos. PHY11-02511 and PHY09-69456.

%

\end{document}